\documentclass[prb,preprint]{revtex4-1} 

\usepackage{graphicx}
\usepackage{epstopdf}

\usepackage{amsfonts,amssymb,amsmath}

\usepackage{bm}

\usepackage[usenames]{color}
\definecolor{Red}{rgb}{1.0,0.0,0.0}
\definecolor{Green}{rgb}{0.0,1.0,0.0}
\definecolor{Blue}{rgb}{0.0,0.0,1.0}
\definecolor{Pink}{rgb}{1.0,0.0,1.0}
\definecolor{Yellow}{rgb}{1.0,1.0,0.0}

\usepackage[applemac]{inputenc}

\def\IH{{\mathrm i}\hbar}
\def\E{{\mathrm e}}
\def\I{{\mathrm i}}
\def\D{{\mathrm d}}

\def\B{\langle }
\def\K{\rangle }
\def\EdefA{\stackrel{\mbox{\tiny def}}{\,=\,}}

\newcommand{\N}{\mathbb{N}}

\begin{document}

\title{A method for solving gyroscopic equations with operators}

\author{Claude Aslangul}
  \email{aslangul@lptmc.jussieu.fr}
\affiliation{Laboratoire de Physique Th\'{e}orique de la Mati\`ere 
Condens\'ee, \\Laboratoire associ\'{e} au
CNRS (UMR 7500),  Sorbonne Universit\'es, 2 place Jussieu, 75252 Paris Cedex 05, France}

\date{\today}

\begin{abstract}

The dynamics of a set of identical spins interacting with another one through a time-dependent coupling gives rise to a gyroscopic equation with a variable Larmor frequency and, more importantly, with an {\it operator} playing the role a Larmor vector. The subsequent technical complexity is due to non-trivial algebraic relations between multiple inner products coming from the non-commutative algebra of the angular momenta. A general formalism is derived giving the integrated solution valid for all values of the involved spins, and several applications of the formalism are treated in details. Among other results, it is seen that, starting from a fully polarised state for the set of identical spins, their total spin can at most only {\it partially} flip (in the mean); this somewhat surprising fact means that the memory of the initial state is kept for ever but varying the coupling constant allows to adjust at will the possible polarisation of the final state. The robustness of the initial state is shown to depend on the nature fermionic or bosonic of the perturbing spin and also on the size of the collection of identical spins.

 \end{abstract}


\maketitle


\section{Introduction}\label{intro}

Due to its specific non commutative algebra, quantum theory yields differential equations which cannot be integrated by using ordinary methods of classical mathematical analy\-sis. The most celebrated example is probably the Schr\"odinger equation with a hamiltonian which does not commute with itself at different times; although formal integration is indeed possible, it requires the introduction of a chronological operator \cite{davydov} ${\cal T}$ in order to preserve the basic algebra of operators.

Another example is provided by the Heisenberg equations, which are rather appealing when one is interested in the dynamics of a few observables only and provide a clear physical meaning since they usually bear strong formal ressemblance with the corresponding classical equations. Nevertheless such an analogy should not be misleading because operators, not scalars, are at the heart of these equations ; in addition, great care must be exercised in handling them since they do not come under the ordinary treatment based on the diagonalization of the relevant dynamical matrix. 

In the following, I will focus on a gyroscopic equation arising within a simple model describing the interaction of a set of $N$ identical discernable spins $\vec S_i$, each of them having a transient interac\-tion with another one $\vec I$; more precisely, the hamiltonian of such a system is taken as : 
\begin{equation}
H(t)=\hbar^{-1}\omega(t)\big(\sum_{i=1}^N\vec S_i\big).\vec I\equiv \hbar^{-1}\omega(t)\,\vec S.\vec I\label{Hamilt}\enspace;
\end{equation}
and is the simplest rotationally invariant hamiltonian for such a system ; for physical reasons, $\omega(t)$ is a fonction vanishing at $t=\pm\infty$ and assumed to be integrable. It assumed that all these spins commute together, namely $[\vec S_i,\,\vec S_j]=0$ for all $i\neq j$ and $[\vec S_i,\,\vec I\,]=0$. 

With $N=2$, the two $\vec S_i$ can be viewed as projectiles thrown into a solid containing a localized impurity; alternatively, if $\vec I$ is a nuclear spin and the $\vec S_i$ are electrons, the above hamiltonian modelizes a temporary effective hyperfine interaction. Anyhow, the following treatment is independent of the character fermionic or bosonic of all the spins, as well as of the number $N$, the total spin $\vec S$ assuming the values $S=-NS_i,\,-NS_i+1,..., +NS_i$; clearly, $\vec S^{\,2}$ is a constant of motion, as well as $\vec I^{\,2}$.

\section{Heisenberg equations and formal integration}\label{HeisenEqn}

Denoting the Heisenberg representation with a roman H as subscript ($\vec S_{\mathrm H}(t)$, etc), the Heisenberg equations are readily written as follows :
\begin{equation}
\frac{\D}{\D t}\vec S_{{\mathrm H}}=\hbar^{-1}\omega(t)(\vec I_{\mathrm H}\times\vec S_{\mathrm H})\enspace,\qquad
\frac{\D}{\D t}\vec I_{{\mathrm H}}=\hbar^{-1}\omega(t)(\vec S_{\mathrm H}\times\vec I_{\mathrm H})\enspace.
\end{equation}

Due to $[S_u,\,I_v]=0$ $\forall\,u,\,v=x,\,y,\,z$, one has $\vec S\times\vec I=-\vec I\times\vec S$, an equality which is invariant in the Heisenberg picture since the latter proceeds through a unitary transformation. The total angular momentum $\vec J=\vec S+\vec I$ is thus a constant of motion, a consequence of the fact that the Hamiltonian Eq.~(\ref{Hamilt}) is rotationally invariant. Thus, one has $\vec J_{\mathrm H}(t)=\vec J$ at all times and, provided the initial state is an eigenvector of $J_z$, $J_z|\Psi(-\infty)\K=M_J\hbar|\Psi(-\infty)\K$, the dynamics is confined to this given eigensubspace of $J_z$, the quantum number $M_J$ being fixed once for all. The elimination of $\vec I_{\mathrm H}$ gives a closed equation for $\vec S_{{\mathrm H}}$ which,
using the commutation relation $\vec S\times \vec S=\IH\vec S$, can be recast in the form:
\begin{equation}
\frac{\D}{\D t}\vec S_{\mathrm H}(t)+\I\omega(t)\vec S_{\mathrm H}(t)=\hbar^{-1}\omega(t)\vec J\times\vec S_{\mathrm H}(t)\enspace;\label{eqngyroSSS}
\end{equation}
the second term in the left-hand side would be absent with classical vectors but the quantum nature of the problem by far does not reduce to this, as shown below. Note that although $\vec S$ and $\vec J$ obey a non-commutative algebra, some equalities are here and there useful in the following; for instance $\vec J.\vec S=
\vec S.\vec J$, $\vec S.(\vec S\times\vec J\,)=(\vec J\times\vec S\,).\vec S=\IH \vec J.\vec S=\vec J.(\vec J\times\vec S\,)=(\vec S\times\vec J\,).\vec J$, 
all of them being still true in the Heisenberg picture. In addition, $\vec J$ being a constant of motion, one has $\frac{\D}{\D t}(\vec J.\vec S_{\mathrm H})\!=\!\vec J.\frac{\D}{\D t}\vec S_{\mathrm H}\!=\!-\I\omega(t)[\vec J.\vec S_{\mathrm H}+\I\hbar^{-1}\vec J.(\vec J\times\vec S_{\mathrm H})]
\!=\!0$ so that the scalar operator $\vec J.\vec S_{\mathrm H}(t)$ is also constant in time
, a fact formally reminiscent of what says an ordinary gyroscopic equation. 
%
%

Setting now $\vec S_{\mathrm H}(t)=\E^{-\I\phi(t)}\,\vec \Sigma(t)$, with $\phi(t)\EdefA \int_{-\infty}^t\omega(t')\,\D t'$, the dynamical equation Eq.~(\ref{eqngyroSSS}) yields:
\begin{equation}
\frac{\D}{\D t}\vec\Sigma(t)=\hbar^{-1}\omega(t)\vec J\times\vec\Sigma(t)
\label{eqngyroSigma}
\end{equation}
which is the basic equation to be solved. It looks like a standard gyroscopic equation with a time-dependent Larmor pulsation, but this is not actually the case, aside from the fact that $\vec \Sigma(t)$ is not a vector operator because $[\Sigma_x,\,\Sigma_y]\neq\IH\Sigma_z$ etc,  since the precession takes place around $\vec J$ which is an {\it operator}\,; this peculiarity carries the full complexity  otherwise absent when the latter is a c-vector, as an external magnetic field for instance. 

By successive integrations on each side of eq.\,(\ref{eqngyroSigma}), one can write the formal conver\-gent series (remember that $\omega(t)$ is assumed to be integrable):
\begin{eqnarray}
\vec\Sigma(t)=\vec S+
\frac{1}{\hbar}\int_{-\infty}^t\D t_1\,\omega(t_1)\,\vec J\times\vec S+\frac{1}{\hbar^2}\int_{-\infty}^t\D t_1\,\omega(t_1)\int_{-\infty}^{t_1}\D t_2\,\omega(t_2)\,
\vec J\times(\vec J\times\vec S\,)+\nonumber\\
\frac{1}{\hbar^3}\int_{-\infty}^t\D t_1\,\omega(t_1)\int_{-\infty}^{t_1}\D t_2\,\omega(t_2)
\int_{-\infty}^{t_2}\D t_3\,\omega(t_3)\vec J\times\big(\vec J\times(\vec J\times\vec S\,)\big)+...
\end{eqnarray}
Since the multiple integrals only involve time-dependent quantities which all commute together, the $n^{\mathrm th}$ integral  is just $\frac{1}{n!}\big[\int_{-\infty}^t\omega(t')\,\D t'\,\big]^n\equiv\frac{1}{n!}[\phi(t)]^n$, so that:
\begin{equation}
\vec\Sigma(t)=\sum_{n\in\N}\frac{[\phi(t)]^n}{n!}\vec {\sf P}_n\enspace,\qquad
\vec {\sf P}_n\EdefA \frac{1}{\hbar^n}\vec J\times\big(\vec J(\times...(\vec J\times\vec S\,))\big)\label{SerieSigma}\enspace,
\end{equation}
$\vec {\sf P}_n$ containing $n$ times the factor $\vec J\times$, and $\vec {\sf P}_0\EdefA\vec S$; once the above series is summed up, $\vec S_{\mathrm H}(t)$ follows according to  $\vec S_{\mathrm H}(t)=\E^{-\I\phi(t)}\,\vec\Sigma(t)$.

At this point, the Heisenberg equation for $\vec S_{\mathrm H}(t)$ is formally integrated and the expectation value at time $t$, $\B\vec S\,\K(t)$, is equal  to $\B\Psi(-\infty)|\vec S_{\mathrm H}(t)|\Psi(-\infty)\K$, simply noted $\B\vec S_{\mathrm H}(t)\K_{-\infty}$ in the following; summing up at this point, one has:
$$
\B\vec S\,\K(t)=\E^{-\I\phi(t)}\B\vec \Sigma\,\K(t)\enspace,\qquad\B\vec \Sigma\,\K(t)\EdefA\sum_{n\in\N}\frac{[\phi(t)]^n}{n!}\B\vec {\sf P}_n\K_{-\infty}\enspace.
$$
The question is now to calculate the average value $\B\vec {\sf P}_n\K_{-\infty}$ of the vectorial products in the expansion Eq.~(\ref{SerieSigma}),
 and then to sum up the series to eventually find the average value $\B\vec S\,\K(t)$; as it will be seen below, the  Heisenberg picture is here not at all trivial.

\section{Algebraic relations and explicit integration}\label{algebraic}

The difficulty lies in the occurence of multiple inner products implying vectorial {\it opera\-tors}\,; this is already seen when one considers the single product; indeed:
$$\hbar \vec{\sf P}_1\equiv\vec J\times\vec S=(\vec S+\vec I\,)\times\vec S=\IH\vec S+\vec I\times\vec S
\enspace;
$$
since $\vec S$ and $\vec I$ commute one has $\vec J\times\vec S+\vec S\times\vec J=2\IH \vec S\neq\vec 0$ (note that the RHS, being pro\-portional to $\hbar$, vanishes in the classical limit, as it must).

Without surprise, things become still more involved for the double vectorial product; for classical vectors, one knows that:
\begin{equation}
\vec U\times(\vec V\times \vec W)=(\vec U.\vec W)\,\vec V-(\vec U.\vec V)\,\vec W\enspace;\label{dblepdvecclas}
\end{equation}
here, the component along ${\mathrm O}x$ of $\vec J\times(\vec J\times \vec S\,)$ is:
\begin{equation}
J_y(J_xS_y-J_yS_x)-J_z(J_zS_x-J_xS_z)=J_yJ_xS_y+
J_zJ_xS_z-(J_y^2+J_z^2)S_x\enspace\label{dblepdvecqu}
\end{equation}
instead of:
$$J_yS_yJ_x+J_zS_zJ_x-(J_y^2+J_z^2)S_x\equiv(\vec J.\vec S\,)J_x-\vec J^{\,2}S_x$$
 deduced from the equality Eq.~(\ref{dblepdvecclas}). The RHS of Eq.~(\ref{dblepdvecqu}) can be written as:
$$J_y(S_yJ_x+\IH S_z)+J_z(S_zJ_x-\IH S_y)+J_x^2S_x-\vec J^{\,2}S_x\enspace;
$$
due to $[J_x,\,S_x]=0$, the second to last term on the right is equal to $J_xS_xJ_x$ so that 
$$\big(\vec J\times(\vec J\times \vec S\,)\big)_x=(\vec J.\vec S\,)J_x+\IH(\vec J\times\vec S\,)_x-\vec J^{\,2}S_x
$$ 
and more generally:
\begin{equation}
\hbar^2\vec{\sf P}_2\equiv\vec J\times(\vec J\times \vec S\,)=(\vec J.\vec S\,)\vec J-\vec J^{\,2}\vec S+\IH(\vec J\times\vec S\,)\enspace;
\label{decompP2}
\end{equation}
as for the single vectorial product, the additional term vanishes when $\hbar=0$.

In order to get insight into the multiple vectorial product of any order, it is useful to calculate $\vec{\sf P}_3$ and $\vec{\sf P}_4$ explicitly; after some straightforward calculation, one finds:
$$
\hbar^3\vec{\sf P}_3\equiv\vec J\times\big(\vec J\times(\vec J\times \vec S\,)\big)=-(\vec J^{\,2}+\hbar^2)\vec J\times \vec S+2\I\hbar(\vec J.\vec S\,)\vec J-\I\hbar\vec J^{\,2}\vec S\enspace,
$$
$$\hbar^4\vec{\sf P}_4=-(\vec J^{\,2}+3\hbar^2)(\vec J.\vec S\,)\vec J+\vec J^{\,2}(\vec J^{\,2}+\hbar^2)\vec S-
\I\hbar(2\vec J^{\,2}+\hbar^2)\vec J\times \vec S\,;
$$
again, all the ``non-classical" terms disappear in the limit $\hbar\rightarrow0$ since:
\begin{equation*}
\vec U\times(\vec U\times(\vec U\times \vec V))=-\vec U^{\,2}\,\vec U\times\vec V\enspace,\quad
\vec U\times(\vec U\times(\vec U\times(\vec U\times \vec V)))=-\vec U^{\,2}\,(\vec U.\vec V)\,\vec U+\vec U^4\,\vec V\enspace.
\end{equation*}
%
%
%
%

These preliminary calculations allow to be convinced that the set $(\vec J.\vec S\,)\vec J, \vec S$ and $\vec J\times \vec S$ is a closed algebra, each of its elements having a clear geometrical meaning; this enables to write the product of any order $\vec{\sf P}_n$ as a linear combination of these three vectorial operators with coefficients which are some functions of the operator $\hbar^{-2}\vec J^{\,2}\EdefA j$; this leads to set:
$$\vec {\sf P}_n=\hbar^{-2}{\bf a}_n(j)(\vec J.\vec S\,)\vec J+{\bf b}_n(j)\vec S+\hbar^{-1}{\bf c}_n(j)\vec J\times \vec S\,\,(n\ge 0)\,,
$$
where the operators ${\bf a}_n$, ${\bf b}_n$ et ${\bf c}_n$ are dimensionless functions of $j$ to be determined and  assumed to commute with $\vec J$, an assumption which will be checked below. The correctness of such statements is confirmed by the existence of a recurrence relationship having a definite solution, obtained by handling the equality for $\vec {\sf P}_{n+1}=\hbar^{-1}\vec J\times \vec {\sf P}_n$:
\begin{equation*}\hbar^{-2}{\bf a}_{n+1}(\vec J.\vec S\,)\vec J+{\bf b}_{n+1}\vec S+\hbar^{-1}{\bf c}_{n+1}\vec J\times \vec S=
\hbar^{-1}\vec J\times\big[\hbar^{-2}{\bf a}_n(\vec J.\vec S\,)\vec J+{\bf b}_n\vec S+\hbar^{-1}{\bf c}_n\vec J\times \vec S\,\big]\enspace.
\end{equation*}
%
%
%
Then, taking Eq.~(\ref{decompP2}) and $[\vec J,\,\vec J.\vec S\,]=0$ into account, the RHS can be recast as:
$$\hbar^{-2}(\I {\bf a}_n+{\bf c}_n)(\vec J.\vec S\,)\vec J-{\bf c}_n\hbar^{-2}\vec J^{\,2}\vec S+\hbar^{-1}({\bf b}_n+\I {\bf c}_n)\vec J\times \vec S\enspace;
$$
from this, one obtains the following linear recurrence between the unknown operators:
\begin{equation}
{\bf a}_{n+1}(j)=\I {\bf a}_n(j)+{\bf c}_n(j)\enspace,\quad
{\bf b}_{n+1}(j)=-j\,{\bf c}_n(j)\enspace,\quad
{\bf c}_{n+1}(j)={\bf b}_n(j)+\I {\bf c}_n(j)\enspace.
\label{recuroperabc}
\end{equation}
%
%
The  initial condition is found by considering $\vec {\sf P}_0=\vec S$ and yields:
\begin{equation}
{\bf a}_0(j)={\bf 0}\enspace,\qquad {\bf b}_0(j)={\bf 1}\enspace,\qquad {\bf c}_0(j)={\bf 0}\enspace.
\label{condinitRePn}
\end{equation}
Such a recursive relation does have a unique solution, a fact which {\it a posteriori} proves that the algebra is indeed closed. In addition, ${\bf a}_0$, ${\bf b}_0$ and ${\bf c}_0$ trivially commute with $\vec J$ and therefore with $\vec J^{\,2}$; due to the recursive relations Eqs.~(\ref{recuroperabc}), this holds true for any value of $n$,  showing that the starting assumption is indeed correct. Note that some of the coefficients of this multidimensional recursion between operators are themselves operators so that the formal solution is not very useful.  For sure, one can obviously write:
$$
\left[\begin{array}{c}{\bf a}_{n+1}\\{\bf b}_{n+1} \\{\bf c}_{n+1}\end{array}\right]={\bf M}\left[\begin{array}{c}{\bf a}_n\\{\bf b}_{n} \\{\bf c}_{n}\end{array}\right]
\enspace,\qquad
{\bf M}\EdefA\left[\begin{array}{ccc} \I & 0 & 1\\0 & 0 & -j \\0 &  1 & \I\end{array}\right]\enspace,
\label{recuPPPn}
$$
so that ${\bf[}{\bf a}_{n}\,{\bf b}_{n}\,{\bf c}_{n}{\bf ]}^\dagger={\bf [}{\bf 0}\,{\bf 1}\,{\bf 0}{\bf ]}^\dagger {\bf M}^n$. The $n^{\mathrm th}$ power of ${\bf M}$ can be calculated by first writing down a characteristic equation to obtain the ``eigenvalues" $\I$, $\frac{\I}{2}[1\pm(1+4j)^{1/2}]$ but  before to go further, one should first define properly the square root of $1+4\hbar^{-2}\vec J^{\,2}$ (which is not impossible) and then achieve the diagonalisation; this route is certainly not the simplest  way to proceed (all the more since  $\vec J^{\,2}$ can have a zero eigenvalue) and it is better to translate all the preceding equalities in terms of scalars, which is now performed. 

Gathering the previous  definitions, the expectation value of $\vec\Sigma$ reads:
\begin{equation}
\B\vec \Sigma\,\K(t)=\hbar^{-2}\B {\bf A}(j,\,t)(\vec J.\vec S\,)\vec J\,\K_{-\infty}+\B {\bf B}(j,\,t)\vec S\,\K_{-\infty}+\nonumber
\\\hbar^{-1}\B {\bf C}(j,\,t)\vec J\times \vec S\,\K_{-\infty}\enspace,\quad
\label{sigmasomme}
\end{equation}
with 
\begin{equation}
\left[\begin{array}{c}{\bf A}(j,t)\\{\bf B}(j,t) \\{\bf C}(j,t)\end{array}\right]\EdefA\sum_{n\in\N}\frac{[\phi(t)]^n}{n!}\left[\begin{array}{c}{\bf a}_n(j)\\ {\bf b}_n(j) \\{\bf c}_n(j)\end{array}\right]
\enspace,
\end{equation}
the question being now to find the relevant matrix elements of the operators ${\bf a}_n$, ${\bf b}_n$ and ${\bf c}_n$. Since the latter are functions of $j\equiv\hbar^{-2}\vec J^{\,2}$ only, it is convenient to introduce the eigenvectors of the total angular momentum, $|\psi_{JM_J}\K$, ($\vec J^{\,2}|\psi_{JM_J}\K=J(J+1)\hbar^2|\psi_{JM_J}\K$, $|S-I|\le J\le S+I$, $ J_z|\psi_{JM_J}\K=M_J\hbar|\psi_{JM_J}\K$, $-J\le M_J\le+J$), which can be obtained by following the standard procedures relevant in the addition of angular momenta \cite{messiah1}$^{,}$ \cite{merzbacher}$^{,}$ \cite{ClAslangulMQ2}. 

With the assumption that the initial state $|\Psi(-\infty)\K$ is an eigenvector of $J_z$  (but not necessarily of $\vec J^{\,2}$) with the given eigenvalue $M_J\hbar$, only the eigenvectors $|\psi_{JM_J}\K$ appear in its expansion, namely $ |\Psi(-\infty)\K=\sum_{J}c_J|\psi_{JM_J}\K$, the number $J$ varying by one unit at a time within the above mentioned interval; in the case where a given value of $J$ can be obtained in several ways, a second label will be required, {\it e.g.} $c_{Ji}$, but this complication is omitted here for simplicity. Accor\-ding to (\ref{sigmasomme}), one now has to calculate the average values $\hbar^{-2}\B {\bf A}(j,\,t)(\vec J.\vec S\,)\vec J\,\K_{-\infty}$, etc;
by letting now the operators ${\bf a}_n(j)$, ${\bf b}_n(j)$ and ${\bf c}_n(j)$ act on the {\it bra}, the following quantities appear:
\begin{equation}
{\sf A}(t)\EdefA\sum_{J}c^*_J\B \psi_{JM_J}|\frac{1}{\hbar^2}(\vec J.\vec S\,)\vec J\,|\Psi(-\infty)\K \sum_{n=0}^{+\infty}\frac{[\phi(t)]^n}{n!}\alpha_n(J)\,,\nonumber
\end{equation}
\begin{equation}
{\sf B}(t)\EdefA\sum_{J}c^*_J\B \psi_{JM_J}|\vec S\,|\Psi(-\infty)\K \sum_{n=0}^{+\infty}\frac{[\phi(t)]^n}{n!}\beta_n(J)\enspace,\nonumber
\end{equation}
\begin{equation}
{\sf C}(t)\EdefA\sum_{J}c^*_J\B \psi_{JM_J}|\hbar^{-1}\vec J\times\vec S\,|\Psi(-\infty)\K \sum_{n=0}^{+\infty}\frac{[\phi(t)]^n}{n!}\gamma_n(J)\enspace,\nonumber
\end{equation}
the numerical functions $\alpha_n$, $\beta_n$ and $\gamma_n$ depending on the quantum number $J$ and result from the replacement of the operator $j$  by its eigenvalue $J(J+1)$ in the operators ${\bf a}_n(j)$, ${\bf b}_n(j)$ and ${\bf c}_n(j)$ respectively; the average value $\B\vec \Sigma\K(t)$ is the sum ${\sf A}(t)+{\sf B}(t)+{\sf C}(t)$. 

Going back now to the relations Eq.~(\ref{recuroperabc}), multiplication on the left by $\B\psi_{J0}|$ and on the right by $|\Psi(-\infty)\K$ yields the following recursion between the scalars $\alpha_n$, $\beta_n$ and $\gamma_n$ which, in matrix form, reads:
$$
\left[\begin{array}{c}\alpha_{n+1}\\\beta_{n+1} \\\gamma_{n+1}\end{array}\right]=M\left[\begin{array}{c}\alpha_n\\ \beta_{n} \\\gamma_{n}\end{array}\right]
\enspace,\quad
M\EdefA\left[\begin{array}{ccc} \I & 0 & 1\\0 & 0 & -J(J+1) \\0 &  1 & \I\end{array}\right]\enspace,
\label{recuPPn}
$$
with the initial condition $\beta_0(J)=1$, $\alpha_0(J)=\gamma_0(J)=0$ (see Eq.~(\ref{condinitRePn})). The matrix $M$ is diagonalizable and can be written as $M=PM_DP^{-1}$ with (temporarily assuming $J\neq0$):
$$
P=\left[\begin{array}{ccc}\vspace{2pt}1 &\frac{\I}{J+1} & -\frac{\I}{J}\\0 & -\I(J+1) & \I J  \\\vspace{2pt}0 & 1 & 1\end{array}\right] 
\enspace,\qquad
P^{-1}=\left[\begin{array}{ccc} 1 & \frac{1}{J(J+1)} &\frac{\I}{J(J+1)} \\0 & \frac{\I}{2J+1} &\frac{J}{2J+1}\\ 0 & -\frac{\I}{2J+1} & \frac{J+1}{2J+1}\end{array}\right]
$$
and:
$$M_D=\left[\begin{array}{ccc} \I & 0 & 0 \\0 &-\I J & 0 \\0 & 0 & \I(J+1) \end{array}\right]\equiv
\left[\begin{array}{ccc} \lambda_0 &0 &0\\0 &\lambda_-& 0 \\0 & 0 &\lambda_+ \end{array}\right]
\enspace.
$$
Note that the above quantities do not have a simple expression in terms of the single variable $J(J+1)$, a reminder of the fact that, in its first version, the recurrence involves operators, not scalars. Anyway, the solution can now be written down:
$$\left[\begin{array}{c}\alpha_n(J)\\\beta_{n}(J) \\\gamma_{n}(J)\end{array}\right]=PM^{n}_D\,P^{-1}\left[\begin{array}{c}0\\1\\0\end{array}\right]\enspace,$$
so that:
\begin{equation*}
\alpha_n(J)=\frac{(2J+1)\lambda_0^n-J\lambda_-^n-(J+1)\lambda_+^n}{J(J+1)(2J+1)}\enspace,
\quad \beta_n(J)=\frac{1}{2J+1}[(J+1)\lambda^n_-+J\lambda^n_+]
\end{equation*}
\begin{equation*}
\gamma_n(J)=\frac{\I}{2J+1}(\lambda^n_--\lambda^n_+)\enspace.
\end{equation*}
It is easily checked that all this agrees with $\vec{\sf P}_n$, $0\le n\le4$, directly  obtained at the beginning. Now, 
%
%
recognizing exponential series in the RHS and inserting all this in the sum ${\sf A}(t)+{\sf B}(t)+{\sf C}(t)$, one gets the final closed explicit expression for $\B\vec S\,\K(t)=\E^{-\I\phi(t)}\B\vec \Sigma\K(t)$  
which eventually yields the desired result for the expectation value of $\vec S$:
\begin{equation}
\B\vec S\,\K(t)=\sum_{J}\frac{c^*_J}{2J+1}\B \psi_{JM_J}|{\cal \vec S}(J,\,t)|\Psi(-\infty)\K\label{Sfinal}
\end{equation}
with:
\begin{equation*}
{\cal \vec S}(J,\,t)\EdefA\hbar^{-2}s_0(J,\,t)(\vec J.\vec S\,)\vec J+s_\parallel(J,\,t)\vec S+
\I\hbar^{-1}s_\perp(J,\,t)\vec J\times\vec S\label{SSfinalexpl}\enspace,
\end{equation*}
%
where the various scalar functions are:
\begin{equation*}
s_0(J,\,t)\EdefA \frac{(2J+1)-J\,\E^{-\I(J+1)\phi(t)}-(J+1)\,\E^{\I J\phi(t)}}{J(J+1)}\enspace,
\end{equation*}
\begin{equation*}
s_\parallel(J,\,t)\EdefA (J+1)\,\E^{-\I(J+1)\phi(t)}+J\,\E^{\I J\phi(t)}\enspace,\quad s_\perp(J,\,t)\EdefA \E^{-\I(J+1)\phi(t)}-\E^{\I J\phi(t)}\enspace.
\end{equation*}
Note that $s_0(J,\,t)$ has the same finite value when the two formal limits $J\rightarrow0_\pm$ are taken and can thus be defined by continuity without any trouble for $J=0$. For $t=-\infty$ ($\phi=0$), all this reduces to $\sum_{J}c^*_J\B \psi_{JM_J}|\vec S\,|\Psi(-\infty)\K
=\B\Psi(-\infty)|\vec S\,|\Psi(-\infty)\K$
as it must. Taking into account the fact that the coefficients of the exponentials are all real\cite{operatorherm} and that $\B \vec S\,\K(t)$ is also real, it is clear that various compensations intervene in the calculation of the different quantities in Eq.~(\ref{Sfinal}) to cancel some complex contributions coming from one additive term to the other; this is technically ensured by the fact that in the summation $J$ varies by one unit at a time. Clearly, in the final expression of $\B \vec S\,\K(t)$, only some angles $J\phi$ and $(J+1)\phi$ do appear and, as a whole, only the real parts of the functions $s_0(J,\,t)$, $s_\parallel(J,\,t)$ and $s_\perp(J,\,t)$ are indeed relevant, allowing to claim that $\B \vec S\,\K(t)$ implies only linear terms of the form $\cos \big(J\phi(t)\big)$ aside from a possible additive real constant.

In other respects, one can anticipate which components of $\B\vec S\,\K$ have a non-zero expectation value. As for the first term $(\vec J.\vec S\,)\vec J$, since $J_x$ and $J_y$ change $M_J$ by $\pm 1$ and since $(\vec J.\vec S\,)$ is a scalar operator, only $(\vec J.\vec S\,)J_z$ can give finite contributions due to the fact that $M_J$ is a good quantum number; the same is true for the second term $\vec S$ since $S_x$ and $S_y$ also change $M_J$ in the same way. Finally, only the component of $\vec J\times\vec S$ along ${\mathrm O}z$ can give non-zero contributions since the two others are linear in $S_x$, $S_y$, $I_x$ and~$I_y$. As a whole, $\B S_x\K(t)$ and $\B S_y\K(t)$ vanish at all times, as contrasted with an ordinary gyroscopic equation for which the two transverse components rotate around the Larmor vector at the Larmor frequency.

Because all these formula only rely on the fundamental algebra of angular momenta, the basic formal result Eq.~(\ref{Sfinal}) is valid in all cases, independently of the actual values of the quantum number $J$, {\it i.e.} of the character (bosonic or fermionic) of the spins, and also of the number of spins $\vec S_i$ adding up to give the angular momentum $\vec S$.

\section{Two examples}\label{Example}

\subsection{{\boldmath$N$} fermions and one boson: {\boldmath$S_i=\frac{1}{2}$, $i=1,\,2,\,...,\,N$}  and {\boldmath$I=1$}}\label{N12121}

As already stated, the general formalism is valid for any number of angular momenta. Still with $I=1$, and assuming that $|\Psi(-\infty)\K$ is the ferro\-magnetic state $|\frac{N}{2}, \frac{N}{2}\,;\,-1\K$, $S$ has the constant value $\frac{N}{2}$ throughout and the dynamics is confined to a three-dimensional subspace, here generated by the vectors, $|\frac{N}{2}\frac{N}{2}\,;\,-1\K$, $|\frac{N}{2}\frac{N}{2}-1\,;\,0\K$ and $|\frac{N}{2}\frac{N}{2}-2\,;\,+1\K$ or equivalently by $\{|\psi_{J\frac{N}{2}-1}\K\}_{J=\frac{N}{2},\,\frac{N}{2}\pm1}$; the list of possible exponentials with these $J$ values shows that one  expects to find $\frac{N}{2}\phi$ and $(\frac{N}{2}+1)\phi$ only in the oscillating terms of $\B S_z\K(t)$.

Using standard formulas of angular momentum theory, one easily writes the following decompositions relating the sets of eigenvectors $|SM\,;\,m\K$ and~$|\psi_{JM_J}\K$:
\begin{eqnarray}
|\psi_{\frac{N}{2}+1\frac{N}{2}-1}\K=\frac{1}{\sqrt{2(N+1)(N+2)}}\Big[\sqrt{2N(N-1)}\,|\frac{N}{2}\frac{N}{2}-2\,;\,+1\K+2\sqrt{2N}\,|\frac{N}{2}\frac{N}{2}-1\,;\,0\K+\nonumber\\
2|\frac{N}{2}\frac{N}{2}\,;\,-1\K\Big]\enspace,\nonumber
\end{eqnarray}
\begin{eqnarray}
|\psi_{\frac{N}{2}\frac{N}{2}-1}\K=\frac{1}{\sqrt{N(N+2)}}\Big[-2\sqrt{N-1}\,|\frac{N}{2}\frac{N}{2}-2\,;\,+1\K+(N-2)|\frac{N}{2}\frac{N}{2}-1\,;\,0\K+\nonumber\\
\sqrt{2N}\,|\frac{N}{2}\frac{N}{2}\,;\,-1\K\Big]
\nonumber\enspace,
\end{eqnarray}
\begin{eqnarray}
|\psi_{\frac{N}{2}-1\frac{N}{2}-1}\K=\frac{1}{\sqrt{N(N+1)}}\Big[\sqrt{2}\,|\frac{N}{2}\frac{N}{2}-2\,;\,+1\K-\sqrt{2(N-1)}\,|\frac{N}{2}\frac{N}{2}-1\,;\,0\K+\nonumber\\
\sqrt{N(N-1)}\,|\frac{N}{2}\frac{N}{2}\,;\,-1\K\Big]\nonumber\enspace;
\end{eqnarray}
\noindent from this, the expression of the initial state is found to be the following:
\begin{equation}
|\Psi(-\infty)\K=\frac{\sqrt{N-1}}{\sqrt{N+1}}|\psi_{\frac{N}{2}-1\frac{N}{2}-1}\K+\frac{\sqrt2}{\sqrt{N+2}}|\psi_{\frac{N}{2}\frac{N}{2}-1}\K+\nonumber\\
\frac{\sqrt2}{\sqrt{(N+1)(N+2)}}|\psi_{\frac{N}{2}+1\frac{N}{2}-1}\K
\nonumber\enspace,
\end{equation}
%
where one can read the coefficients $c_J$ appearing in the general formula Eq.~(\ref{Sfinal}). This being done, and remarking that $(\vec J.\vec S\,)J_z$ does now contribute, some rather tedious but elementary calcula\-tions eventually yield the following expression for the expectation value of $S_z$ at \mbox{time $t$:}
%
\begin{equation}
\hbar^{-1}\B S_z\K(t)=\frac{N}{2}+\frac{4 (N-1)}{ N (N+1)} \Big[ \cos \big[\frac{N}{2}\phi(t)\big]-1\Big]+\frac{8 N }{ (N+1) (N+2)^2}\Big[\cos \big[\big(\frac{N}{2}+1\big) \phi(t)\big]-1\Big]\enspace.\label{SzFinalNfermionsbis}
\end{equation}
%
Note that $\B S_z\K(t)$ never vanishes if $N\ge4$.
%
%
%

\begin{figure}[htbp]
\centerline{\includegraphics[width=200pt]{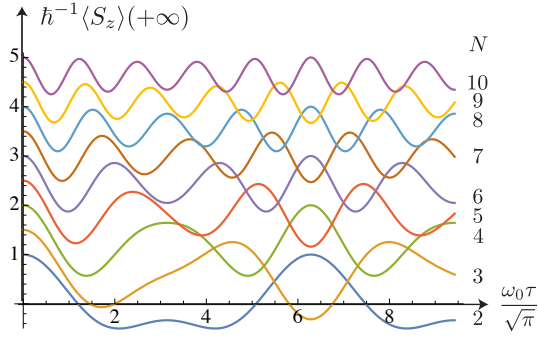}
}
\caption{Final value $\B S_z\K(+\infty)$ as a function of the dimensionless coupling parameter $\omega_0\tau$.}
\label{AJPArticleJJSBosonNSpins12Fin}
\end{figure}

For $N\gg1$, $\B S_z\K(t)$ has the approximate expression:
\begin{equation*}
\hbar^{-1}\B S_z\K(t)\simeq\frac{N}{2}-\frac{8}{N}\sin^2\frac{N}{4}\phi(t)-\hspace{50pt}
\\\frac{8}{N^2}\Big[\cos\frac{N}{2}\phi(t)-\cos \big(\frac{N}{2}+1\big) \phi(t)\Big]+{\cal O}(N^{-3})\,,
\end{equation*}
showing that $\lim_{N\rightarrow\infty}\B S_z\K(t)=\frac{N}{2}\hbar$ for all times and that the lower bound of $\B S_z\K(t)$ is then close to $\hbar(\frac{N}{2}-\frac{8}{N})$. More generally, $\B S_z\K$ is greater than the function of $N$ where the two cosine are simultaneously equal to $-1$ (only possible when $N$ is odd), so that:
$$\hbar^{-1}\B S_z\K\ge\frac{N^4+4 N^3-12 N^2-64 N+64}{2 N (N+2)^2}\EdefA b(N)\stackrel{N\gg1}{=}\frac{N}{2}-\frac{8}{N}+{\cal O}(N^{-3})\enspace;
$$
In the classical limit $N\rightarrow\infty$, $\hbar\rightarrow0$, $N\hbar={\mathrm C}^{\mathrm st}\EdefA2S_{\mathrm{cl}}$, one has $\B S_z\K(t)=S_{\mathrm {cl}}\,\,\forall\,t$ as it must.

\begin{figure}[htbp]
\centerline{\includegraphics[width=390pt]
{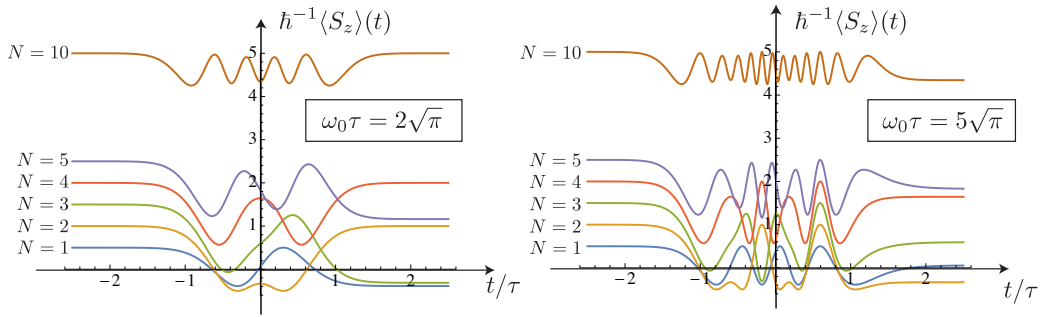}}
\caption{For $N$ fermions $S_i=\frac{1}{2}$ and one boson $I=1$, time variation of the average value $\B S_z\K(t)$ for a gaussian interaction and for two values of the dimensionless coupling parameter $\omega_0\tau$.}
\label{ArticleJJSBosonSzN121}
\end{figure}

The final value of $\B S_z\K$ is a function of the strength of the coupling between spins; for definiteness, let us choose a gaussian perturbation, $\omega(t)=\omega_0\,\E^{-(t/\tau)^2}$, for which one has\cite{GradhsteinRyzhik}  $\phi(t)=\frac{\sqrt{\pi}}{2}\omega_0\tau[1+\Phi(\frac{t}{\tau})]$. Setting $\phi(+\infty)=\sqrt\pi\,\omega_0\tau$ in the expression (\ref{SzFinalNfermionsbis}), one obtains the Fig.~\ref{AJPArticleJJSBosonNSpins12Fin}; although $\B S_z\K(+\infty)$ is never strictly equal to its initial value, this is approximately the case, for large $N$, when $\omega_0\tau=\frac{4}{N}k\sqrt\pi$, $k$ integer.
The variation in time of $\B S_z\K(t)$ with the gaussian perturbation is shown in  Fig.~\ref{ArticleJJSBosonSzN121} in a few cases.

All this demonstrates that the set of fermions is, as $N$ increases, more and more robust against the bosonic perturbation, as is stronger the memory effect since for any strength of the coupling the final expectation of the fermion and boson spins along ${\mathrm O}z$ are closer and closer to their  initial values. This effect is still enhanced when the perturbation is modulated at a high frequency.

\subsection{{\boldmath$N+1$} fermions: {\boldmath$S_i=I=\frac{1}{2}$}}\label{121212}

\begin{figure}[htbp]
\centerline{\includegraphics[width=220pt]{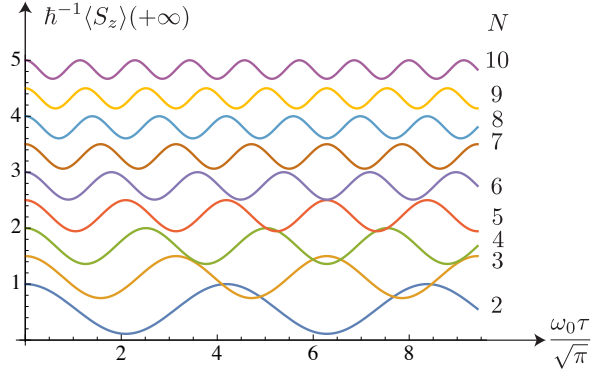}}
\caption{For $N$ spins $S_i=\frac{1}{2}$ and with $I=\frac{1}{2}$: final value of $\B S_z\K$ as a function of the dimensionless coupling parameter $\omega_0\tau$.}
\label{ArticleJJSBosonN1212SzFin}
\end{figure}

The case of $N+1$ fermions, namely $S_i=I=\frac{1}{2}$,   
is treated along the same lines and I just quote the results; as before $\B S_x\K(t)=0=\B S_y\K(t)$ and in addition only an oscillation at $\frac{N+1}{2}\phi$ can be forecast. The final result is :
\begin{equation}
\hbar^{-1}\B S_z\K(t)=\frac{N}{2}+\frac{2N}{(N+1)^2}\Big[\cos\big[\frac{N+1}{2}\phi(t)\big]-1\Big]\nonumber\enspace,
\label{SzFinal121212N}
\end{equation}
showing that the lower bound is now $\frac{N(N-1)(N+3))}{2(N+1)^2}\stackrel{N\gg 1}{=}\frac{N}{2}-\frac{4}{N}+\frac{8}{N^2}+{\cal O}(N^{-3})$
quite close to but slightly greater than $b(N)$ obtained in the previous case $I=1$; again, the classical situation is recovered when the proper limit is performed. Note that $\B S_z\K(t)$ now never vanishes as soon as $N\ge2$.
%
%
The final value of $\B S_z\K$ is plotted in Fig.~\ref{ArticleJJSBosonN1212SzFin} as  a function of $\omega_0\tau$.

\begin{figure}[htbp]
\centerline{\includegraphics[width=390pt]{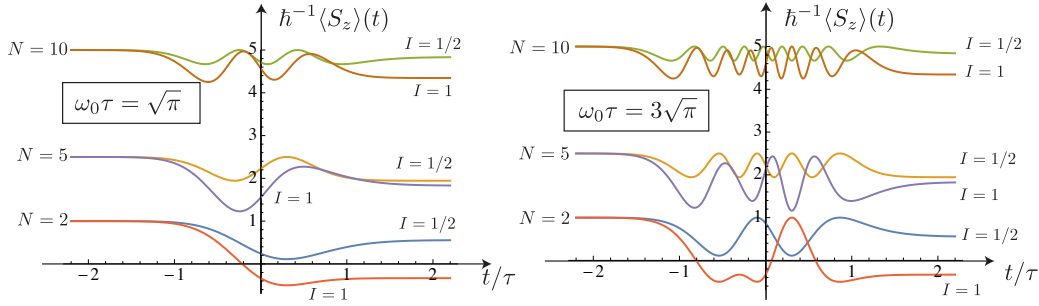}
}
\caption{For $N$ spins $S_i=\frac{1}{2}$ and with $I=\frac{1}{2},\,1$: time variation of the average value $\B S_z\K(t)$ for a gaussian interaction and for two values of the dimensionless coupling parameter $\omega_0\tau$.}
\label{ArticleJJSBosonSzRecapitul}
\end{figure}

For the comparison, Fig.~\ref{ArticleJJSBosonSzRecapitul} displays the differences for the time variation of $\B S_z\K(t)$ between the two cases $I=\frac{1}{2}$ and $I=1$. In particular, if the memory effect is always greater for the former value of the perturbing spin $I$,  the ratio recedes  while oscillating and goes to $1$ as $N$ increases: for a large aggregate, it is hard to find out whether the perturbing spin is a fermion or a boson. Also note that the classical limit is again correctly recovered.

%

\section{Summary and conclusions}\label{Conclusions}

A general formalism has been given for finding the solution of a gyroscopic equation characterised by the fact that the Larmor vector around which precession occurs is an {\it operator}, as contrasted with the ordinary case. This implies a somewhat involved procedure coming from the non-commutative algebra of the various implied angular momenta. This formalism is valid for bosons and fermions and for any number of spins. Several applications have been given, describing the transient interaction of an aggregate of $N$ fermions starting in its ferromagnetic state with an incoming particle, either a boson or a fermion. It was shown that, on the average, none of the spins $\vec S$ ans $\vec I$ can be fully reversed by the collision, witnessing of a kind of memory effect since the final state always keeps track of the initial one. This effect is more pronounced when the perturbing spin is a fermion but the difference is evanescent when the size of the aggregate increases. The probability to find a reversed spin is bounded above and this bound decreases when $N$ increases, going to zero in the infinite $N$ limit. Finally, analogies and differences with an ordinary gyroscopic equation have been stressed.


\begin{thebibliography} {99}
\bibitem{NotMatrix}Note that these scalar produits are not matrix elements since {\it bras} and {\it kets} do not belong to the same orthonormalized basis.
\end{thebibliography}
\end{document}